\documentclass[twocolumn,showpacs,preprintnumbers,amsmath,amssymb]{revtex4}

\usepackage{graphicx}
\usepackage{dcolumn}
\usepackage{bm}
\usepackage{amsmath}
\usepackage{SIunits} 

\begin{document}


\title{Robust \emph{ab initio} calculation of condensed matter: transparent
  convergence through semicardinal multiresolution analysis}

\author{I.P. Daykov}
\affiliation{Laboratory of Atomic and Solid State Physics, Cornell University, Ithaca, NY 14853}

\author{T.D. Engeness}
\affiliation{Department of Physics, Massachusetts Institute of Technology, Cambridge, MA 02139}

\author{T.A. Arias}
\affiliation{Laboratory of Atomic and Solid State Physics, Cornell University, Ithaca, NY 14853}

\date{DRAFT: \today}

%
%
\begin{abstract}
We present the first wavelet-based all-electron density-functional
calculations to include gradient corrections and the first in a solid.
Direct comparison shows this approach to be unique in providing
systematic ``transparent'' convergence, convergence with \emph{a
priori} prediction of errors, to beyond chemical (millihartree)
accuracy. The method is ideal for exploration of materials under novel
conditions where there is little experience with how traditional
methods perform and for the development and use of chemically accurate
density functionals, which demand reliable access to such precision.
\end{abstract}

\pacs{71.15.Ap, 71.15.Mb, 71.15.Nc}

\keywords{wavelets,multiresolution analysis, \emph{ab initio},density
functional theory,electronic structure,chemical accuracy,MgO}

\maketitle

Over the last several decades, the \emph{ab initio} density-functional
approach, which replaces the many-body wave function with a set of
single-particle ``Kohn-Sham'' orbitals moving in an effective
potential\cite{KS}, has opened to first-principles study a diverse
array of condensed matter phenomena ranging from plasticity, diffusion
and surface reconstruction to melting and chemical
reactions\cite{bible}.  While density-functional theory is exact in
principle, the true form of the effective potential as a functional of
the orbitals is unknown.  Current research is pushing available
approximations to this functional toward \emph{chemical
  accuracy}\cite{refs5-6fromBurke}, typically defined as predicting
bond-breaking energies to 1~kcal/mol = 1.6~millihartree.  The prime
motivation for this precision is the ability to predict rates of
microscopic processes at room temperature, where the relevant energy
scale is $k_{\mathrm B} T\approx$~1~millihartree.

Unfortunately, current feasible representations for the Kohn-Sham
orbitals, such as the plane-wave pseudopotential approach\cite{bible}
or the atomic sphere methods\cite{lapwlmto}, are not easily improved
systematically to millihartree accuracy.  For instance, although
plane-wave bases converge systematically with basis size,
pseudopotential calculations require significant experience in the
construction of reliable potentials\cite{bible}.  And, although atomic
sphere methods use the Coulomb potential of the nuclei directly, they
remain also an art involving many parameters and requiring significant
expertise to ensure millihartree accuracy consistently\cite{lapwbook}.
As testimony to these difficulties, it is not uncommon to find in the
literature disagreements over the predictions of density-functional
theory.

To illustrate this, Table~\ref{tbl:results} presents the lattice
constant, cohesive energy and bulk modulus of a very simple solid, MgO
in its rock-salt structure, both as measured in
experiments\cite{crc,Ecoh-nscatt,Ecoh-interf} \emph{and} and as
calculated within the local spin-density approximation (LSDA) when
using various representations for the Kohn-Sham
orbitals\cite{taurian,schromberger,mehl,cohen84,cappellini}.  No two
methods in the table agree consistently on the predictions of LSDA.
For each property, the spread in predictions is on the order of the
discrepancy from the experiment.  Without direct estimation of the
errors, it is difficult to judge how much of each discrepancy is
inherent in LSDA or due to biases built into the different
representations.  Hence, the use of existing methods is questionable
when exploring materials under novel conditions where there is little
experience with how such methods perform, as in the study of materials
at geological pressures under which atomic cores begin to overlap.
(There is recent interest in \emph{ab initio} study of MgO under such
conditions\cite{science}.)  Finally, lack of consistent access to
millihartree precision hampers development and eventual use of
chemically accurate functionals.
                
\begin{table}
\begin{tabular}{r|lcc}
 & a [$\angstrom$] & E$_{\mathrm coh}$ [eV] & B [\mega bar]\\\hline\hline
Experiment\cite{crc,Ecoh-nscatt,Ecoh-interf} & 4.21 & 10.33 & 1.55 -- 1.62\\ \hline
LMTO\cite{taurian}   & 4.09  & 10.67 & 1.71 \\
FP-LMTO\cite{schromberger}& 4.16  &  & \\
FP-LAPW\cite{mehl}& 4.167 & & 1.72 \\
pseudopotential\cite{cohen84} & 4.191 & $\;\;9.96 (+0.5)$ & 1.46\\
pseudopotential\cite{cappellini} & 4.125 & 11.80 & 1.56
\end{tabular}
\caption{Structural properties of MgO
}
\label{tbl:results}
\end{table}

The purpose of this letter is to demonstrate for the first time that a
new general representation, a wavelet-like multiresolution analysis,
provides for solid-state electronic structure calculations an
unprecedented level of precision and a new capability for
\emph{transparent convergence}, systematic convergence with an
extremely simple and predictable scaling for the errors.  We also
demonstrate the first wavelet calculations to employ generalized
gradient approximations (GGAs).  We find GGAs to fit seamlessly within
our approach, without special concerns such as the
discontinuities at the sphere boundaries which arise in the atomic
sphere methods\cite{singh91}.

To date, the application of multiresolution analysis to \emph{ab
initio} calculations has been limited solely to very simple systems,
such as single-electron atoms\cite{prl}, the diatomic hydrogen
molecule\cite{tymczak}, diatomic oxygen using
pseudopotentials\cite{chou}, purely electrostatic problems without
electronic structure\cite{goedecker}, or all-electron calculations of
atoms\cite{mgras}.  Only recently have all-electron wavelet
calculations of small molecules appeared in the
literature\cite{cho,torkel_molecules}.  Here, new
techniques\cite{ross,rmp} enable us to present the first such
calculations in solids and the first to include gradient corrected
density functionals.

\emph{Multiresolution analysis ---} A multiresolution analysis
consists of a basis set of spatially localized functions which
describe fluctuations on a hierarchy of length scales, each separated
by a factor of two, with the basis functions representing each of
these levels of resolution organized on simple rectilinear grids.
(Ref.~\cite{rmp} provides a detailed review.)  The central, nontrivial
mathematical result of multiresolution analysis is that, with
appropriately chosen basis functions, this multilevel description is
mathematically equivalent to a uniform grid of basis functions on the
finest level of resolution\cite{Mallat,Meyer}.  This key result allows
for \emph{a priori} knowledge of convergence.  In particular, the use
of third-order interpolating basis functions in the present work
implies that all errors scale as the fourth power of the spacing on
the finest level, a factor of sixteen for each additional level of
resolution.

The superiority of the multiresolution analysis over the uniform
representation comes from the unique way in which the analysis
represents information.  Because fine-scale coefficients carry
information only about high spatial frequencies, they drop rapidly to
zero with distance away from the nuclei\cite{cho,rmp}.  Thus,
\emph{restriction} of the basis, elimination of functions from finer
levels far from the nuclei, has controllably negligible effect on the
outcome of the calculation.  In practice, sufficient functions can be
restricted from the basis so that the basis size and workload grow
linearly with the number of levels of resolution, making convergence
exponential with basis size\cite{cho,mgras}.

The primary reason for the limitation of wavelet calculations to very
simple systems in the past has been the lack of efficient algorithms
suited to the solution of non-linear partial differential equations.
We recently introduced new algorithms\cite{ross,rmp} which are faster
than the approaches used in the older wavelet electronic-structure
works by some three to four orders of magnitude.  With these methods,
all-electron calculations now require an effort of the same order of
magnitude as their pseudopotential
counterparts\cite{torkel_molecules}, opening the possibility of
all-electron wavelet calculations of non-trivial systems.

Given a multiresolution representation, our density-functional
calculations proceed by straightforward expansion of the Kohn-Sham
Lagrangian\cite{rmp,dftpp} or, equivalently, energy
functional\cite{KS}, employing nuclear potentials constructed as
described in \cite{torkel_molecules} and employing either the Vosko,
Wilk and Nussair parameterization (VWN) of 
LSDA\cite{lsda-vwn,lsda-vwn2} or the Perdew, Burke and Ernzerhof
parameterization (PBE96) of GGA\cite{PBE96}.  The calculations then
locate the stationary point of the functional using standard
preconditioned conjugate gradient methods with analytic continuation
as in \cite{analcontprl} to maintain the orthonormality constraints
among the Kohn-Sham orbitals.

Refs.~\cite{rmp,torkel_molecules} give the full details of our
implementation, with the exception of four extensions which proved
critical in carrying out calculations in solids to high precision: (1)
treatment of boundary conditions for Bloch states, (2) symmetrization
of the electron density, (3) expansion of the electron density on
higher resolution grids, and (4) the extension of our approach to
include gradient corrections to the exchange-correlation energy.

(1) In a finite multiresolution basis, the common practice of
expanding the periodic parts $u_{nk}(r)$, rather than the full Bloch
orbitals $\psi_{nk}(r)=e^{ikr} u_{nk}(r)$, does not ensure
\emph{extensivity}, that decreasing cell size while correspondingly
increasing Brillouin-zone sampling results in identical total
energies.  However, because wavelet bases share the translational
symmetries of the lattice, their expansion coefficients satisfy a
discrete Bloch's theorem, $c_{x+R}=e^{ikR} c_x$, where $c_x$ and $R$
are, respectively, the expansion coefficient for the basis function
centered at $x$ and any lattice vector.  We thus implement Bloch
states by storing the coefficients $c_x$ for each orbital $\psi_{nk}$
in a single representative cell and producing all needed coefficients
$c_{x+R}$ by multiplying the corresponding coefficient $c_x$ by
$e^{ikR}$ (``twisted boundary conditions'').

\def\cI{{\mathcal I}}
\def\cJ{{\mathcal J}}
\def\cO{{\mathcal O}}
\def\cD{{\mathcal D}}

(2) For k-point sampling, we use the scheme of Monkhorst and Pack
\cite{monkhpack} with symmetry folding of the Brillioun zone.  For
this folding to be exact, the symmetrized electron density must be
expandable in a multiresolution basis respecting the symmetries of the
crystal.  Accordingly, we symmetrize the density with the operator $s
\equiv \cI P \cJ S \cI P \cJ$, composed from the usual real-space
symmetrization operator $S$, a projector onto basis functions which
have all symmetry partners $P$, and the forward $\cI$ and inverse
$\cJ$ wavelet transforms of \cite{rmp}.  Moreover, differentiation of
the total energy shows that the self-consistent potential requires a
further symmetrization with the operator $s^\dagger$, a different
operator from $s$ because $\cJ \ne \cI^\dagger$ for non-orthogonal bases.

(3) In our older approach, which samples real-space quantities only at
the centers of basis functions\cite{torkel_molecules}, errors in
approximate evaluation of the total energy dominate errors from
incomplete representation of the orbitals.  Accordingly, we now save
considerable computational effort by expanding the orbitals in a
basis of exactly half the spatial resolution of that representing the
other quantities.  This proves critical to the present calculations
and requires the introduction of new transform operators to be
described in depth in a forthcoming publication.

(4) Gradient corrections implement simply and naturally within our
framework.  Using the notation established in \cite{rmp},
the exchange-correlation energy
$$
E_{xc}=\int d^3x\,\, n(x) \epsilon_{xc} \left(\partial_{x_1}
n(x),\partial_{x_2} n(x),\partial_{x_3} n(x),n(x) \right)
$$
becomes
\begin{equation} \label{eq:ExcGGA}
E_{xc}=\left(\cJ n\right)^\dagger \cO \cJ \epsilon_{xc}(\cD_1 \cJ
n,\cD_2 \cJ n,\cD_3 \cJ n,n),
\end{equation}
where $n$ is a vector of values of the electron density at sampling
points at the centers of the basis functions, the $\cD_i$ are matrices
of values of the $\partial_{x_i}$ derivative of each basis function at
each such sample point, and $\cO$ is the matrix of overlaps among
basis functions.  The exchange-correlation potential then follows
directly by differentiating (\ref{eq:ExcGGA}) with respect to $n$,
which introduces no new operators other than the Hermitian conjugates
of the $\cD_i$.

\emph{Transparent convergence ---} To illustrate
the feasibility of larger calculations in complex systems, we study an
eight atom cubic supercell of MgO.  Without pseudizing or treating
differently the core states or core regions, our results converge
directly to the true LSDA/GGA predictions as a function of only three
parameters: number of iterations in the self-consistent solution of
the Kohn-Sham equations, the number of k-points in the sampling of the
Brillouin zone, and the size of the multiresolution basis.  The
remaining paragraphs of this section address these three parameters
one by one.

Despite the broad range of length scales in the calculation, the
convergence of conjugate-gradient minimization is extremely good (0.12
digits/iteration), even when compared to that of plane-wave
pseudopotential calculations.  As \cite{rmp,torkel_beyond} discuss,
this results from the use of our specific non-orthogonal basis with a
simple diagonal preconditioner.  This approach achieves millihartree
precision (corresponding to six significant figures) within just forty
iterations of starting from randomized atomic wave functions.  We
typically ran eighty iterations.

Exploring convergence with respect to Brillioun zone sampling (with a
multiresolution basis somewhat smaller than that which the
calculations below employ) we find that that eight k-points, which
reduce to one special point at $[0.25\, 0.25\, 0.25]$ in the
irreducible wedge, suffice to converge the energy to 0.3 millihartree
per chemical unit, better than chemical accuracy.

\begin{figure}
\begin{center}
\scalebox{0.32}{\includegraphics{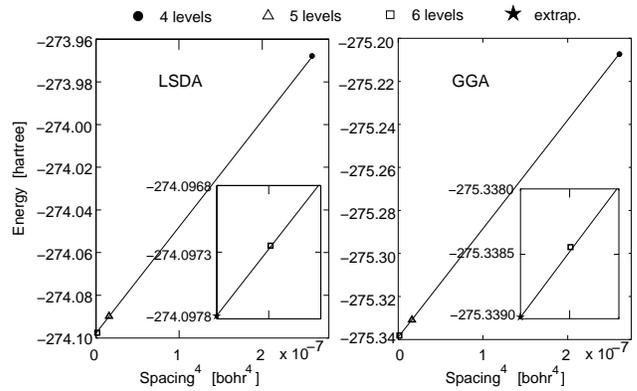}}
\end{center}
\caption{Convergence of total energy per MgO chemical unit with basis
size: LSDA (left panel), GGA (right panel).  Insets are comparable in
size to square symbols in main plots.}
\label{fig:gridconv}
\end{figure}

A key, novel result of this work is the simple form of the convergence
to the full all-electron result with increasing basis set and the high
precisions which this allows us to reach.  Figure~\ref{fig:gridconv}
shows for both LSDA and GGA the total energy per MgO chemical unit as
a function of the fourth power of the spacing on the finest level for
the multiresolution grids in Table~\ref{tbl:MgOgridstruct} when
truncated at four, five and six levels of refinement.  The results
demonstrate that the basis convergence of GGA is almost identical that
that of LDA, with no adverse effects from the presence of second
derivatives of the density in the exchange-correlation potential.  The
data clearly exhibit the \emph{a priori} expected quartic convergence,
thus demonstrating that the restriction of the basis has negligible
effect and that the calculation has entered the asymptotic convergence
regime where there are no hidden convergence ``shoulders''.  The
quality of the linear fit empowers extrapolation to infinite
resolution (stars in the figure insets) with an error of only
$\sim$14~\emph{micro}hartree, far below the tolerances described for the most
accurate of the standard approaches\cite{mehl}.  We also see that,
even without extrapolation, six levels of refinement suffice to give
the energy to within 0.5 millihartree per chemical unit.  Thus, the
present approach gives simple, transparent knowledge of the precision
of the calculation at any stage.  We believe the simplicity of the
convergence of multiresolution analysis to the all-electron result in
a practical calculation to be unique in the field of electronic
structure.

\begin{table}
\begin{center}
\vspace{0.1cm}
\begin{tabular}{c||c|c}
Level &  \multicolumn{2}{c}{${N_x,N_y,N_z}$} \\ \hline \hline
0     & \multicolumn{2}{c}{44, 44, 44}  \\\hline \hline
& Mg & O \\\hline \hline
1     &\ 24, 24, 24 \    & \  24, 24, 24 \   \\\hline
2     &24, 24, 24    &  24, 24, 24   \\\hline
3     &24, 24, 24    &  28, 28, 28   \\\hline
4     &28, 28, 28    &  28, 28, 28   \\\hline
5     &32, 32, 32    &  32, 32, 32
\end{tabular}
\end{center}
\caption{Restriction employed in the present work:
dimension (${N_x,N_y,N_z}$) of the cubic grids of basis functions
on the top coarse periodic level (0), and lower levels (1-5) centered
on magnesium and oxygen nuclei.}
\label{tbl:MgOgridstruct}
\end{table}

\emph{Results ---} Table~\ref{tbl:results-new} summarizes the
predictions of the LSDA-VWN and GGA-PBE96 parameterizations of
density-functional theory, when computed with one special k-point and the
six-level restriction of the multiresolution analysis from
Table~\ref{tbl:MgOgridstruct}, which the discussion above establishes
to give millihartree precision.  Note that, to allow direct comparison
with the values available in the literature, the table lists the value
of the binding energy at the {\em experimental} lattice constant.
Comparison of Table~\ref{tbl:results} with Table~\ref{tbl:results-new}
shows that, of the traditional methods, only FP-LAPW consistently
reproduces the fully converged LSDA results to chemical accuracy.
Further, it is now possible to make a definitive determination of the
relative transferability among pseudopotentials.  Finally, we can now
judge unambiguously the predictions of LSDA and GGA for MgO.  As
generally the case for LSDA, the lattice constant is within a few
percent (-1.5\%) of the experiment, the system is slightly over bound
(+14\%), and the bulk modulus is in error by several percent (5-10\%,
depending upon to which experiment we compare).  For GGA relative to
LSDA, we find the typical expansion of the lattice constant,
correction of the over-binding (when including atomic
spin-polarization energies) and decrease in the bulk modulus.  For
MgO, GGA improves these results dramatically.

\begin{table}
\begin{tabular}{r|lll}
 & a [$\angstrom$] & E$_{\mathrm coh}$ [eV] & B [\mega bar]\\\hline\hline
LSDA  & 4.161$\pm$0.003 & 11.80$\pm$0.03&  1.71$\pm$0.03\\
GGA  & 4.221$\pm$0.004 & 11.90$\pm$0.03&  1.64$\pm$0.03\\
\multicolumn{2}{c}{\ } & \multicolumn{1}{c}{10.4$^\dagger$} &  \\
\end{tabular}\\
$^\dagger$Including estimate for spin-polarization energy of
atomic oxygen based on LSDA/LDA difference of 1.5~eV.
\caption{Multiresolution analysis calculation of structural properties
of MgO within LSDA and GGA.}
\label{tbl:results-new}
\end{table}

\emph{Conclusions ---} We report the first solid-state all-electron
calculations within a wavelet-like multiresolution analysis and the
first multiresolution calculations to include gradient corrections.
The results give the first \emph{independent} verification that, of
the traditional methods, FP-LAPW is by far the most accurate and gives
the first unambiguous demonstration that GGA outperforms LSDA for MgO.
We also demonstrate multiresolution analysis to be unique among
electronic-structure representations in exhibiting extremely simple
and predictable convergence to far beyond millihartree accuracy.  This
approach, therefore, is ideal for novel problems such as the study of
the impact of core polarization effects on electron energy loss and
X-ray absorption spectra and the investigation of materials under
geological pressures, applications where the use of the traditional,
less systematic approaches remains more an art than a science.

\begin{acknowledgments}
This work supported in part by the NSF Information Technology
Research program (Award No. 0085969) and in part by the US DOE ASCI
ASAP Level~2 program (Contract No. B347887).  TDE would like to
thank the Research Council of Norway.  IPD would like to acknowledge
support from the Cornell Department of Physics.  Computational
support by the Cornell Center for Materials Research
supported under NSF DMR-9632275.
\end{acknowledgments}


\bibliography{prl}

\end{document}